\documentclass[aps,pra,twocolumn,superscriptaddress]{revtex4}
\usepackage{amsfonts,amssymb,graphicx}
\usepackage{amsmath}
\usepackage[usenames,dvipsnames]{xcolor}
\usepackage{stackrel}
\usepackage[normalem]{ulem}
\usepackage{color}
\begin{document}

\title{Two-exciton bound states in the Merrifield exciton model}
\author{Ning Wu}
\email{wunwyz@gmail.com}
\affiliation{Center for Quantum Technology Research, and Key Laboratory of Advanced Optoelectronic Quantum Architecture and Measurements (MOE), School of Physics, Beijing Institute of Technology, Beijing 100081, China}
\author{Runtian Li}
\affiliation{Center for Quantum Technology Research, and Key Laboratory of Advanced Optoelectronic Quantum Architecture and Measurements (MOE), School of Physics, Beijing Institute of Technology, Beijing 100081, China}
\author{Dazhi Xu}
\email{dzxu@bit.edu.cn}
\affiliation{Center for Quantum Technology Research, and Key Laboratory of Advanced Optoelectronic Quantum Architecture and Measurements (MOE), School of Physics, Beijing Institute of Technology, Beijing 100081, China}
\begin{abstract}
Understanding the formation and dynamics of two-exciton bound states (BSs) is crucial for harnessing multi-exciton phenomena and designing next-generation organic semiconductor materials. While various mechanisms such as intermolecular exciton-exciton interactions or multi-level intramolecular configurations can generate these BSs, a fundamental understanding of their underlying lattice dynamics remains a key challenge. Within this context, the Merrifield exciton model stands out as a premier framework for capturing the intricate quantum mixing between single-site doubly-excited states and two-particle scattering continuum. In this work, by constructing a set of exact two-exciton Bloch states and applying a plane-wave ansatz to treat the resulting inhomogeneous tridiagonal matrices, we provide a complete, rigorous solution to the two-exciton problem in a finite-size Merrifield chain. We analytically map the parameter regions that support two-exciton BSs, proving that at most two distinct types of BSs can emerge, located either above or below the scattering continuum. Furthermore, we provide a quantitative interpretation of how these two types of BSs evolve as a function of the total wave number within the whole Brillouin zone. The obtained closed-form expressions for the eigenenergies and wave functions offer valuable benchmarks for exploring exciton dynamics and optical responses in related molecular aggregates.
\end{abstract}

\maketitle
\section{Introduction}
\par Organic semiconductor materials and molecular aggregates have attracted intense research interest over the past few decades due to their widespread applications in next-generation optoelectronic devices~\cite{CR2007,CR2018}. Unlike their inorganic counterparts where delocalized Wannier excitons are relevant, organic materials are characterized by highly localized Frenkel excitons~\cite{Frenkel1931}. At high excitation densities or within specific molecular packing geometries, the Frenkel excitons can interact with each other giving rise to multi-exciton phenomena. Among these, the formation of two-exciton bound states (BSs) (also referred to as biexcitons) is of paramount importance~\cite{Merrifield,Agranovich1968,Mukamel,Juzeliunas1997,Juzeliunas1998,Agranovich}. These BSs serve as critical intermediate pathways in advanced, non-linear optoelectronic mechanisms such as singlet exciton fission~\cite{CR2010,Casanova} and triplet-triplet annihilation/fusion~\cite{TTA1963,TTA1968,TTA2010}, both of which offer a theoretical pathway to surpass the classic Shockley-Queisser efficiency limit~\cite{SQ} in solar energy conversion.
\par The well-known Frenkel exciton model serves as a paradigm for energy transport in molecular assemblies by treating the system as a collection of interacting two-level molecules, each consisting of a ground state $|g\rangle$ and a single excited state $|e\rangle$~\cite{Frenkel1931}. However, excitons are forced to be hard-core bosons (or spins-1/2) that cannot occupy the same site~\cite{Suna1963,Spano1991PRL}, which forbids the formation of two-exciton BSs. The microscopic mechanisms responsible for trapping two independent excitons into a co-propagating quasiparticle generally fall into two distinct physical categories. The first relies on spatial inter-site interactions, where attractive nearest-neighbor Coulomb forces overcome the kinetic energy of exciton hopping and pulls the two particles together~\cite{Agranovich1968,Mukamel,Juzeliunas1997,Agranovich}. The second category originates from local intramolecular electronic structures~\cite{Merrifield,Juzeliunas1998,Spano1995PRL,MayJCP,JCP2015}. In this scenario, individual molecules possess multiple excited energy levels, allowing a single site to host a doubly-excited state. When the energy of this single-site double excitation resonance closely matches the combined energy of two individual and separate excitons, a localized quantum mixing occurs. This intramolecular coupling acts as an effective local potential well that is capable of binding two wandering excitons even in the absence of explicit attractive exciton-exciton interactions.
\par  To fundamentally describe this latter category of multi-level intramolecular mixing, Merrifield proposed a seminal tight-binding lattice model in 1959 to study the interactions of excitation waves in a one-dimensional molecular crystal~\cite{Merrifield}. Besides the excited state $|e\rangle$, Merrifield introduces a second higher-lying excited state $|f\rangle$. He finds that two-exciton BSs can emerge even in the absence of inter-site attractive forces, provided the exciton fusion (fission) process $|ee\rangle\to|fg\rangle, |gf\rangle$ ($|fg\rangle, |gf\rangle\to |ee\rangle$) occurs with sufficient intensity. Despite the model's age, it remains a foundational theoretical benchmark in modern organic electronics, directly mapping onto the intermediate states observed in singlet fission materials~\cite{MayJCP,JCP2015}. In addition, a two-exciton BS in the usual description corresponds to an eigenstate in which two nearest-neighbor excitons bound to each other and propagate together through the molecular crystal, which resembles much a two-magnon exchange BS with two spin derivations mainly located on two nearest-neighboring sites that long observed in quantum spin systems~\cite{Bethe,Wortis,Silberglitt,Tonegawa}.
\par Using the ``center-of-mass" coordinates, Merrifield successfully converts the two-exciton problem into a single-particle one and identifies the parameter regions supporting different numbers of two-exciton BSs. However, Merrifield solved the two-exciton problem mainly in the thermodynamic limit, and hence only identified a single type of BS with a monotonically decreasing wave function. More importantly, the evolution of the nature of the BSs with varying wave number in the  Brillouin zone is not clear yet. Recently, a new theoretical method that combines a set of exact two-magnon Bloch states and a plane-wave ansatz was proposed to solve the two-magnon problem in higher-spin XXZ and $J_1$-$J_2$ chains~\cite{PRB2024_1,PRBL2024,physcp2025}. We find that the two-exciton problem in the Merrifield model shares similar mathematical structures as the two-magnon problem in higher-spin spin chains with single-ion anisotropy.
\par In this work, we use the abovementioned method to completely solve the two-exciton problem in a finite-size Merrifield chain. Two types of two-exciton BSs to which we refer as the f BS and ee BS are revealed, which respectively corresponds to the case that the two excitons mainly occupy the $|f\rangle$ state on a single molecule or the $|e\rangle$ state on two nearest-neighboring molecules. We rigorously prove that these two types of BSs are the only possible ones and they lie either above or below the two-exciton scattering continuum. In the thermodynamic limit, our eigenenergies and eigenstates recover those of Merrifield~\cite{Merrifield}. In addition, we analytically determine parameter regions that support different numbers and types of two-exciton BSs. Although the nature of the BSs show complex features within the Brillouin zone, we are able to quantitatively explain the evolution of the BSs with varying wave number that observed in numerical simulations. The obtained explicit expressions for the eigenenergies and wave functions  of all the scattering and bound states in the finite-size case will play an essential role in the study of exciton dynamics and nonlinear optical responses of molecular systems~\cite{Spano1995PRL,JCP2025}.
\par The rest of the paper is organized as follows. In Sec.~\ref{SecII} we introduce the Merrifield exciton model and block diagonalize it using a set of two-exciton Bloch states. In Sec.~\ref{SecIII} we use the plane-wave ansatz to solve the Bloch Hamiltonian and discuss the resulting scattering and bound states in detail. Sec.~\ref{SecIV} is devoted to a quantitative interpretation of the evolution of the bound states with varying wave number. Conclusions are drawn in Sec.~\ref{SecV}.
\section{The Merrfield exciton model and two-exciton bound states}\label{SecII}
\subsection{Hamiltonian}
\par
\begin{figure}
\includegraphics[width=.51\textwidth]{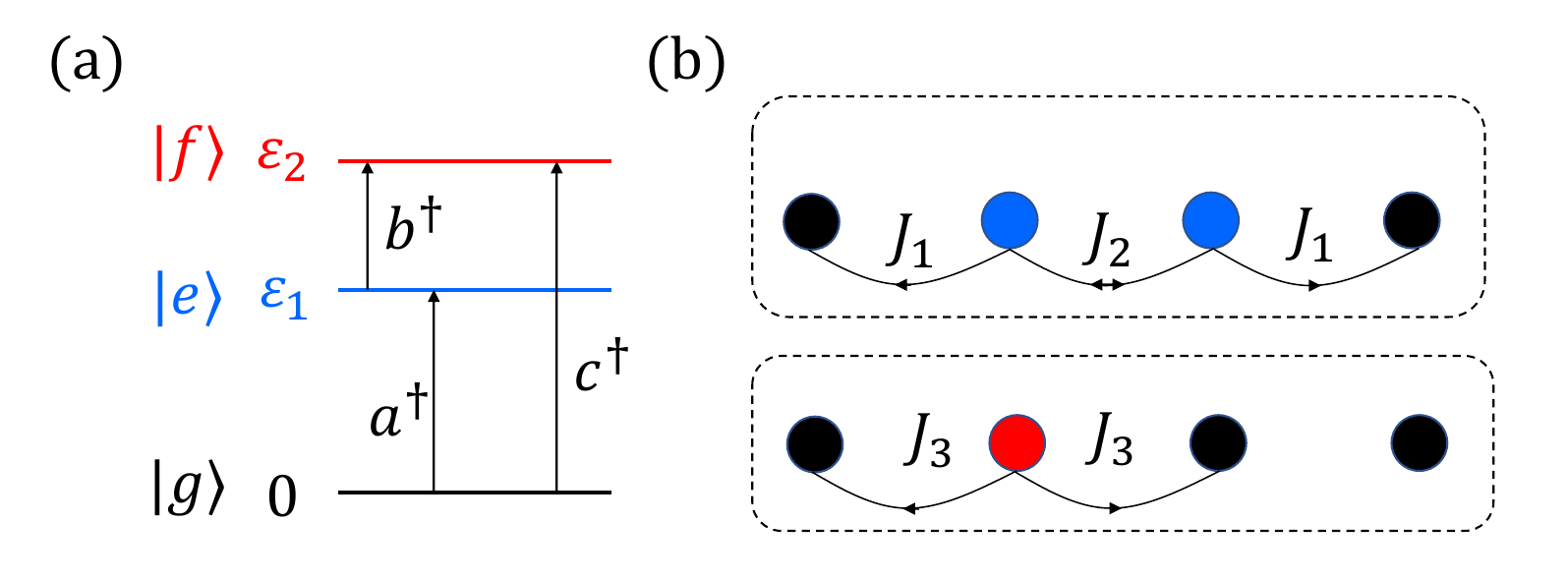}
\caption{(a) Level structure of an individual molecule in the Merrifield exciton model. (b) The three exciton hopping channels measured by $J_1$, $J_2$, and $J_3$. The black, blue, and red solid circles indicate a molecule in the ground state $|g\rangle$, the first excited state $|e\rangle$, and the second excited state $|f\rangle$, respectively.}
\label{Fig1}
\end{figure}

 Consider a finite-size one-dimensional molecular crystal consisting of an array of molecules, in which the ground state $|g\rangle_j$ with energy $0$, the first excited state $|e\rangle_j$ with energy $\varepsilon_1$, and the second excited state $|f\rangle_j$ with energy $\varepsilon_2$ of each molecule are considered [see Fig.~\ref{Fig1}(a) for the level structure of the molecule]. The three states form a complete set on molecule $j$ and are connected by the three raising operators
\begin{eqnarray}
a^\dag_j& \equiv & |e\rangle_j ~_j\langle g|,\nonumber\\
b^\dag_j &\equiv & |f\rangle_j ~_j\langle e|,\nonumber\\
c^\dag_j &\equiv&  |f\rangle_j ~_j\langle g|.
\end{eqnarray}
Note that $b^\dag_j=c^\dag_j a_j$. We say that a molecule in the first (second) excited state $|e\rangle$ ($|f\rangle$) carries one exciton (two excitons). The Hamiltonian for the Merrifield exciton model reads~\cite{Merrifield}
\begin{eqnarray}\label{Hamil}
H&=&H_0+H_1,\nonumber\\
H_0&=&\varepsilon_1\sum^N_{j=1}a^\dag_ja_j+\varepsilon_2\sum^N_{j=1}c^\dag_jc_j,\nonumber\\
H_1&=&J_1\sum^N_{j=1}(a^\dag_ja_{j+1}+a^\dag_{j+1}a_{j})\nonumber\\
&&+J_2\sum^N_{j=1}(b^\dag_ja_{j+1}+a^\dag_{j+1}b_{j}+b^\dag_{j+1}a_{j}+a^\dag_{j}b_{j+1})\nonumber\\
&&+J_3\sum^N_{j=1}(c^\dag_jc_{j+1}+c^\dag_{j+1}c_{j}),
\end{eqnarray}
where $H_0$ is the on-site energy term and $H_1$ describes the exciton hopping between nearest-neighbor molecules. There are three types of exciton hoppings [Fig.~\ref{Fig1}(b)]~\cite{Merrifield}: (1) An exciton can transfer to a nearest-neighbor ground-state molecule with strength $J_1$, corresponding to the usual exciton hopping in the conventional Frenkel exciton model~\cite{Frenkel1931}; (2) For two nearest-neighbor molecules both in their first excited state $|e\rangle$, the exciton on one of the two molecules can transfer to the other with strength $J_2$, leaving the former in its ground state $|g\rangle$ and the latter in the second excited state $|f\rangle$. This process is also named exciton fusion since the two e-excitons are fused into a single f-exciton. However, we avoid using this term because we have defined $|f\rangle$ as a two-exciton state; (3) The two excitons on a molecule in $|f\rangle$ can jointly transfer to a nearest-neighbor ground-state molecule with strength $J_3$. For simplicity, throughout this work we assume that $N$ is even and impose periodic boundary conditions.
\par It is easy to check that the total number of excitons
\begin{eqnarray}
M=\sum^N_{j=1}a^\dag_ja_j+2\sum^N_{j=1}c^\dag_jc_j
\end{eqnarray}
is a conserved quantity, so that $H_0$ can be rewritten as
\begin{eqnarray}
H_0=M\varepsilon_1+(\varepsilon_2-2\varepsilon_1)\sum_j c^\dag_j c_j.
\end{eqnarray}
\par The real-space one-exciton basis states are simply
\begin{eqnarray}
|i\rangle=a^\dag_i|0\rangle,
\end{eqnarray}
where $|0\rangle=|g\ldots g\rangle$ is the state with all the molecules in the ground state. For each wave number $k$ in the set
\begin{eqnarray}
K_0=\left\{-\pi,-\pi+\frac{2\pi}{N},\ldots,0,\ldots,\pi-\frac{2\pi}{N}\right\},
\end{eqnarray}
we can define a one-exciton Bloch state
\begin{eqnarray}
|\xi(k)\rangle=\frac{1}{\sqrt{N}}\sum^{N-1}_{n=0}e^{ikn}T^n|1\rangle,
\end{eqnarray}
where $T$ is the lattice translation operator satisfying $T|i\rangle=|i+1\rangle$. It is easy to check that $|\xi(k)\rangle$ is an eigenstate of $H$ with eigenenergy
\begin{eqnarray}
E_1(k)=\varepsilon_1+2J_1\cos k.
\end{eqnarray}
It is obvious that only the Frenkel hopping with strength $J_1$ takes place in the one-exciton subspace.
\subsection{Physical relevance of the model}
\par It is worth noting that Merrifield's model was proposed in 1959, a time when neither the laser nor the conceptual framework of nonlinear optics had yet emerged. Under weak excitation conditions such as solar irradiation, the probability of having multiple monomers excited or a single monomer in a higher-lying excited state is very small, which made the two-exciton problem appear experimentally inaccessible. Merrifield himself acknowledged that the model was ``too simple to represent any real system accurately"~\cite{Merrifield}, and as a result the work did not receive much attention for decades and its significance is obscured by the absence of the very optical tools that would later make it directly relevant.
\par The prescience of Merrifield's work is now becoming apparent through a series of developments that span nearly six decades. From a theoretical point of view, at least it provides the simplest lattice setting in which two-exciton BSs can emerge from the interplay between on-site multi-level structure and nearest-neighbor hopping, without invoking an explicit attractive potential. In 1995, Knoester and Spano studied the two-photon absorption spectrum of a one-dimensional chain of three-level molecules, which turns out to be equivalent to the Merrifield chain in the case of $J_3=0$~\cite{Spano1995PRL} [their state $|b\rangle$ ($|c\rangle$) corresponds to our state $|e\rangle$ ($|f\rangle$)]. They derive an analytical expression for the two-photon absorption spectrum and find that the presence of the third level $|c\rangle$ can give rise to zero, one, or two additional peaks in the spectrum depending on the system parameters. However, these results are presented within the framework of nonlinear optical response functions and Fermi resonance analogies, without explicit reference to the bound-state picture that Merrifield had employed decades earlier. Nevertheless, the physics Knoester and Spano uncovered is deeply connected to Merrifield's original analysis. 
\par This re-emergence of the two-exciton problem in the 1990s was followed, more recently, by its adoption in studies of exciton-exciton annihilation and singlet fission, where two-exciton states and their interactions occupy a central position. In 2014, May introduced a kinetic theory of exciton-exciton annihilation to revisit the same physical scenario within a modern dynamical framework~\cite{MayJCP}. Notably, May's model also explicitly incorporates two distinct excited states per molecule and most recently it has been used to study coherent biexciton transport in molecular aggregates~\cite{JCP2026}.
\par Actually, the Merrifield model has been explicitly adopted in singlet fission research. Teichen and Eaves proposed a many-body model to describe collective aspects of singlet fission in molecular crystals~\cite{JCP2015}. In the limit where the nearest-neighbor triplet-triplet interaction $\chi$ is neglected, the two-triplet sector of their Hamiltonian reduces to Merrifield's two-exciton problem, with the $|e\rangle$ ($|f\rangle$) state playing the role of a triplet (singlet) excited state. It should emphasized that this correspondence is purely mathematical. In addition, they only discuss (by means of the Bethe ansatz) the formation of two-triplet bound states with increasing $\chi$ in the limit of $J_2\to0$, which corresponds to BSs induced by the attractive interaction rather than the multi-level structure.
\par The mapping between $|e\rangle$ ($|f\rangle$) and a triplet (singlet) suggests that the Merrifield model can be adapted to describe spin states of multiple excitons. Furthermore, as a physically useful and nontrivial model, the Merrifield exciton Hamiltonian is simple enough to admit an analytically treatment. In particular, as we will see in Sec.~\ref{SecIII}, the formation mechanism of two-exciton BSs can be analyzed explicitly across the entire Brillouin zone.
\par Finally, we emphasize that the analysis in this work is restricted to the dissipationless Merrifield model. In real molecular crystals, two-exciton bound states are expected to have a finite lifetime due to exciton-phonon coupling, spin-lattice relaxation, static disorder, and radiative or non-radiative decay. None of these mechanisms is included in the Hamiltonian (\ref{Hamil}), and consequently all eigenstates obtained here are exact stationary eigenstates with infinite lifetime. This idealization is intentional: the goal of this work is to provide a complete and rigorous solution of the idealized lattice model, which can then serve as a basis for incorporating dissipation perturbatively or through a self-energy. 
\subsection{Two-exciton subspace: the Bloch states}
\par Following Ref.~\cite{JCP2015}, in the following we assume
\begin{eqnarray}
\varepsilon_1,~\varepsilon_2,~J_2>0~\mathrm{and}~J_1,~J_3<0.\nonumber
\end{eqnarray}
We choose $J_1=-1$ and set $|J_1|$ as an energy unit. We mainly focus on the two-exciton subspace with $M=2$~\cite{Merrifield}. There are $\binom{N}{2}+N=N(N+1)/2$ real-space two-exciton basis states,
\begin{eqnarray}
|i,j\rangle&=&a^\dag_i a^\dag_j|0\rangle,~ i<j \nonumber\\
|i,i\rangle&=&c^\dag_i |0\rangle.
\end{eqnarray}
\par Following Ref.~\cite{PRB2022}, we introduce for each $k\in K_0$ and $0\leq r<\frac{N}{2}$ the following two-exciton Bloch states
\begin{eqnarray}
|\xi_r(k)\rangle&=&\frac{e^{i\frac{rk}{2}}}{\sqrt{N}}\sum^{N-1}_{n=0}e^{ikn}T^n|1,1+r\rangle,
\end{eqnarray}
where $T$ is the translation operator defined by $T|i,j\rangle=|i+1,j+1\rangle$ and the index $r$ measures the distance between the two local excitons. However, for $r=N/2$ one has to define~\cite{PRB2022}
\begin{eqnarray}
|\xi_{\frac{N}{2}}(k)\rangle&=&e^{i\frac{Nk}{4}}\sqrt{\frac{2}{N}}\sum^{\frac{N}{2}-1}_{n=0}e^{ikn}T^n|1,1+\frac{N}{2}\rangle,
\end{eqnarray}
where $k\in K_1=\{-\pi,-\pi+\frac{4\pi}{N},\ldots,0,\ldots,\pi-\frac{4\pi}{N}\}$ ($k\in K_1=\{-\pi+\frac{2\pi}{N},\ldots,0,\ldots,\pi-\frac{2\pi}{N}\}$) if $N/2$ is even (odd) and the complement of $K_1$ will be denoted as $K'_1$.
\par Straightforward calculations show that for each $k\in K_1$ the matrix representation of $H$ in the ordered basis $\{|\xi_0(k)\rangle,|\xi_1(k)\rangle,\ldots,|\xi_{\frac{N}{2}}(k)\rangle\}$ is
\begin{eqnarray}\label{H2kCk}
\mathcal{H}_2(k)=2\varepsilon_1+C_k h_2(k),
\end{eqnarray}
where $C_k\equiv\cos \frac{k}{2}\geq0$ and
\begin{eqnarray}\label{h2}
h_2(k)=\left(
      \begin{array}{cccccccc}
      A_k &   2J_2   &   & &    &    &   &   \\
          2J_2   & 0 &  -2  &  &   &   &   &   \\
          &  -2   & 0 &  -2    & &   &   &   \\
          &   &  -2   &  0 &  &    &   &   \\
           &   &    &  & \ddots&   &   &   \\
          &   &   &     & &0 &  -2  &   \\
          &   &   &    &  &-2   & 0 &  -2 \sqrt{2}   \\
          &   &   &    & &  &   -2\sqrt{2}  & 0 \\
      \end{array}
    \right),\nonumber\\
\end{eqnarray}
is a $(N/2+1)\times(N/2+1)$ tridiagonal matrix with
\begin{eqnarray}
A_k&\equiv&(\tilde{\varepsilon}+2J_3\cos k)/C_k,\label{Ak}\\
\tilde{\varepsilon}&\equiv&\varepsilon_2-2\varepsilon_1.
\end{eqnarray}
For the boundary mode $k=-\pi$ with $C_k=0$ the matrix $C_kh_2(k)$ should be understood as diagonal,
\begin{eqnarray}\label{kpi}
C_{-\pi}h_2(-\pi)=\mathrm{diag}(\tilde{\varepsilon}-2J_3,0,\ldots,0).
\end{eqnarray}
The above equation shows that the model always has an eigenstate $|\xi_0(-\pi)\rangle$ with energy $\varepsilon_2-2J_3$ and $N/2$ degenerate eigenstates $|\xi_1(-\pi)\rangle$, $|\xi_2(-\pi)\rangle,\ldots,|\xi_{N/2}(-\pi)\rangle$ with energy $2\varepsilon_1$.
\par For each $k\in K'_1$ the matrix representation of $H$ in the ordered basis $\{|\xi_0(k)\rangle,|\xi_1(k)\rangle,\ldots,|\xi_{\frac{N}{2}-1}(k)\rangle\}$ is $\mathcal{H}'_2(k)=2\varepsilon_1+C_k h'_2(k)$, where $h'_2(k)$ is a $\frac{N}{2}\times\frac{N}{2}$ matrix that can be obtained by removing the last column and the last row from $h_2(k)$. The above procedure is equivalent to performing a center-of-mass analysis and converts the two-particle problem into a single-particle one with inhomogeneous effective on-site energies and hoppings. Merrifield also obtained a Bloch matrix that is similar to $h_2(k)$ in the thermodynamics limit $N\to\infty$~\cite{Merrifield}.
\par The $(N/2+1)\times(N/2+1)$ tridiagonal matrix $\mathcal{H}_2(k)$ can be diagonalized to give the eigenenergies $E^{(\alpha)}_2(k)$ and eigenstates $|\psi^{(\alpha)}(k)\rangle$ ($\alpha=1,2,\ldots,N/2+1$). We use the convention
\begin{eqnarray}
E^{(1)}_2(k)\leq E^{(2)}_2(k)\leq\cdots\leq E^{(N/2+1)}_2(k),
\end{eqnarray}
and write
\begin{eqnarray}
|\psi^{(\alpha)}(k)\rangle=\sum^{N/2}_{r=0}V^{(\alpha)}_r(k)|\xi_r(k)\rangle,
\end{eqnarray}
where the wave functions form a column $V^{(\alpha)}\equiv(V^{(\alpha)}_0(k),\ldots,V^{(\alpha)}_{N/2}(k))^{\mathrm{T}}$ that is an eigenvector of the matrix $h_2(k)$.
\subsection{Two-exciton bound states}
\par We now introduce the definition of two-exciton BSs. As we will rigorously prove in the next section, there are possibly two types of two-exciton BSs, to which we refer as the f BS and the ee BS. In analogy to the two types of two-magnon BSs appearing in the higher-spin XXZ chain~\cite{PRBL2024}, an eigenstate $|\psi^{(\alpha)}(k)\rangle$ is called an f BS if the absolute value of the wave functions satisfy
\begin{eqnarray}
|V^{(\alpha)}_0(k)|>|V^{(\alpha)}_1(k)|>\cdots >|V^{(\alpha)}_{N/2}(k)|,
\end{eqnarray}
and is called an ee BS if
\begin{eqnarray}
|V^{(\alpha)}_0(k)|<|V^{(\alpha)}_1(k)|>\cdots >|V^{(\alpha)}_{N/2}(k)|.
\end{eqnarray}
The above definitions indicate that in an f (ee) two-exciton BS the two excitons are mainly located on the same site (two nearest-neighbor sites).
\par Let us first consider two special cases. (1) For $k=-\pi$, it can be seen from Eq.~(\ref{kpi}) that the eigenstate $|\xi_0(-\pi)\rangle$ is an f BS with the two excitons exactly located on a single site and has the lowest (highest) energy if $\tilde{\varepsilon}<2J_3$ ($\tilde{\varepsilon}>2J_3$); (2) When $J_2=0$ the $(1,1)$ element of $h_2(k)$ is decoupled from the remaining $\frac{N}{2}\times\frac{N}{2}$ block and the Bloch state $|\xi_0(k)\rangle$ itself is an f BS for any $k$, which means that the f BS can exist even in the absence of the $|ee\rangle \leftrightarrow|gf\rangle,|fg\rangle$ hopping process. This is not surprising since the second excited state $|f\rangle$ carries two excitons by definition. The $\frac{N}{2}\times\frac{N}{2}$ block describes the conventional Frenkel exciton model that does not support any two-exciton BS unless for example exciton-exciton attraction is introduced~\cite{Agranovich1968,Mukamel,Juzeliunas1997,Agranovich}, which again highlights the role of $J_2$ in the formation the ee BS.
\par Actually, we will show that an ee BS occurs only if $J_2>1$, and the necessary condition for the coexistence of the two types of BSs is $J_2>\sqrt{2}$. We will also prove that if a BS exists it must be either the lowest eigenstate $|\psi^{(1)}(k)\rangle$ or the highest eigenstate $|\psi^{(N/2+1)}(k)\rangle$.
\section{Solution of the two-exciton problem: A plane-wave ansatz}\label{SecIII}
\par In this section, we provide a complete solution to the two-exciton problem by using a plane-wave ansatz. To verify the correctness of the method, we first present the energy spectrum obtained by numerical diagonalization of $h_2(k)$ for various sets of parameters. The observed features of the energy spectrum, especially the evolution of two-exciton BSs with varying wave number $k$, are then interpreted in a transparent way through a plane-wave ansatz solution of $h_2(k)$. We focus on wave numbers in the set $K_1$ as the case of $k\in K'_1$ can be similarly analyzed. We also assume that $N/2$ is even for simplicity.
\subsection{Numerical results}
\begin{figure*}
\includegraphics[width=.82\textwidth]{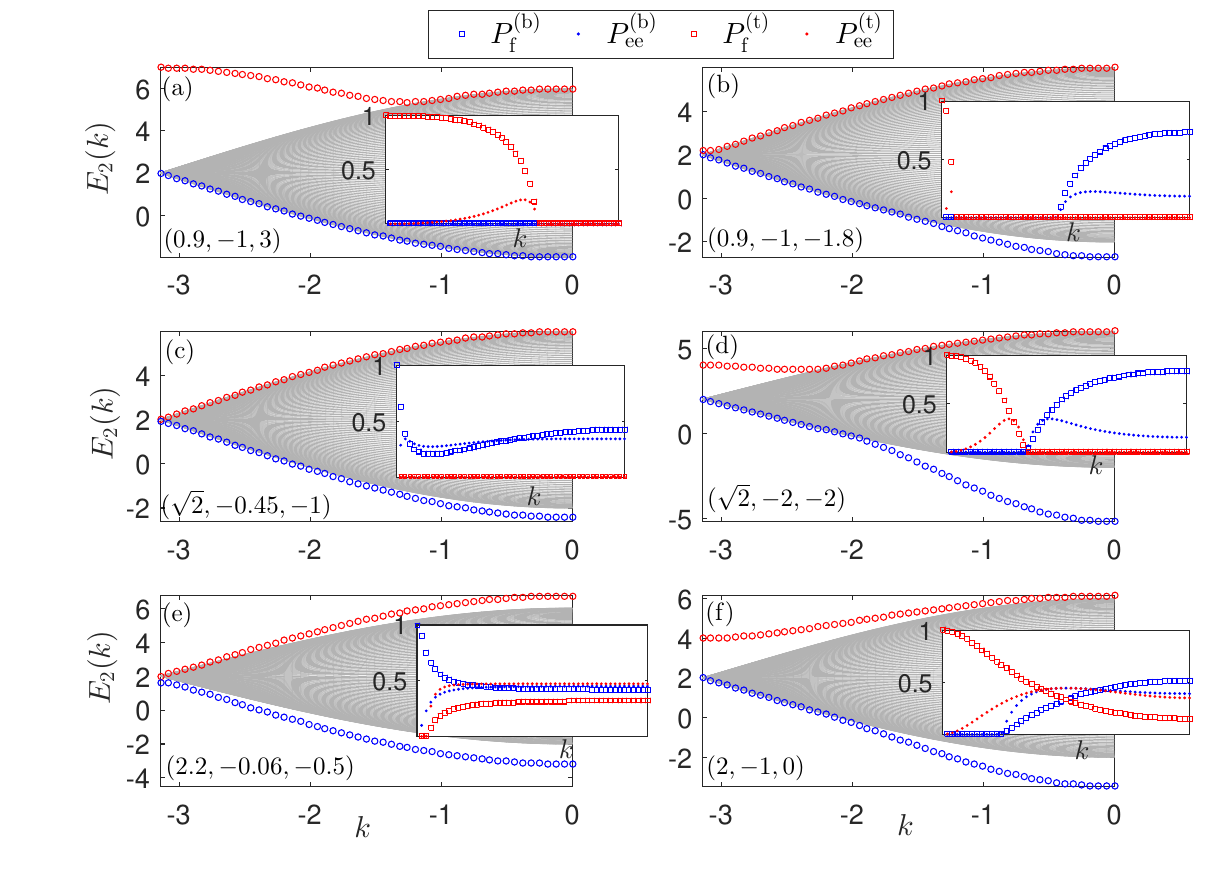}
\caption{Full energy spectra $E_2(k)$ in the two-exciton subspace of the Merrifield exciton model calculated through numerical diagonalization of the Bloch Hamiltonian $\mathcal{H}_2(k)$ (\ref{H2kCk}) for $N=200$, $\varepsilon_1=1$, and $k\in K_1$. The remaining three parameters $(J_2,J_3,\tilde{\varepsilon})$ are shown in the left bottom corner of each panel. The lowest and highest two levels $E^{(1)}_2(k)$ and $E^{(N/2+1)}_2(k)$ are indicated by blue and red circles, respectively. The middle gray region shows the two-exciton continuum. The insets show the evolution of the weights of the two Bloch states $|\xi_0(k)\rangle$ and $|\xi_1(k)\rangle$ in the two states: $P^{(\mathrm{b})}_{\mathrm{f}}(k)=|V^{(1)}_0(k)|^2$ (blue box), $P^{(\mathrm{b})}_{\mathrm{ee}}(k)=|V^{(1)}_1(k)|^2$ (blue dot), $P^{(\mathrm{t})}_{\mathrm{f}}(k)=|V^{(N/2+1)}_0(k)|^2$ (red box), and  $P^{(\mathrm{t})}_{\mathrm{ee}}(k)=|V^{(N/2+1)}_1(k)|^2$ (red dot).} 
\label{numericalfig}
\end{figure*}
\par Figure~\ref{numericalfig} shows the two-exciton energy spectrum of $H$ calculated by numerically diagonalizing the matrix $h_2(k)$ for $N=200$ and $\varepsilon_1=1$. The results for six sets of parameters, $(J_2,J_3,\tilde{\varepsilon})=(0.9,-1,3)$, $(0.9,-1,-1.8)$, $(\sqrt{2},-0.45,-1)$, $(\sqrt{2},-2,-2)$, $(2.2,-0.06,-0.5)$, and $(2,-1,0)$, are presented. As mentioned above, if the BSs exist they must occupy either the lowest (blue circles) or the highest level (red circles). To see the nature of these two states, we plot in the insets of Fig.~\ref{numericalfig} the weights of the two Bloch states $|\xi_0(k)\rangle$ and $|\xi_1(k)\rangle$ in the two states $|\psi^{(1)}(k)\rangle$ and $|\psi^{(N/2+1)}(k)\rangle$, i.e., $P^{(\mathrm{b})}_{\mathrm{f}}(k)=|V^{(1)}_0(k)|^2$, $P^{(\mathrm{b})}_{\mathrm{ee}}(k)=|V^{(1)}_1(k)|^2$, $P^{(\mathrm{t})}_{\mathrm{f}}(k)=|V^{(N/2+1)}_0(k)|^2$, and  $P^{(\mathrm{t})}_{\mathrm{ee}}(k)=|V^{(N/2+1)}_1(k)|^2$.
\par It can be seen that the evolution of the BSs with varying wave number $k$ exhibits complicated features depending on the parameters. When $J_2=0.9<1$ and $J_3=-1$, a high-lying (low-lying) f BS appears near the zone boundary $k=-\pi$ (zone center $k=0$) for $\tilde{\varepsilon}=3$ ($\tilde{\varepsilon}=-1.8$), see Fig.~\ref{numericalfig}(a) [Fig.~\ref{numericalfig}(b)]. In addition, a high-lying f BS also shows up near $k=-\pi$ for $\tilde{\varepsilon}=-1.8$. When $J_2=\sqrt{2}$, the BSs survive throughout the whole zone [Fig.~\ref{numericalfig}(c) and (d)]. For $(J_3,\tilde{\varepsilon})=(-0.45,-1)$, the lowest level corresponds to an f BS near $k=-\pi$, but soon changes to an ee BS as $k$ slightly increases. Interestingly, the f BS is restored at a larger $k$ [Fig.~\ref{numericalfig}(c)]. For $(J_3,\tilde{\varepsilon})=(-2,-2)$, the high-lying (lowest-lying) BS merges into the continuum from the left (right) at the same wave number $k=-2\pi/3$, around which both the two BSs are of ee-type [Fig.~\ref{numericalfig}(d)]. When $J_2$ becomes larger, the two types of BSs can coexist within a wide range of the zone, accompanied by transitions between different types BSs [Fig.~\ref{numericalfig}(e) and (f)].
\par The above numerical results about the energy spectra and the evolution of two-exciton BSs are so complicated that it seems difficult to capture their features in a quantitative way. Compared to the two-magnon problem in a higher-spin XXZ chain where only two parameters (the exchange anisotropy $\Delta$ and the single-ion anisotropy $D$) are relevant~\cite{PRBL2024}, here we have three parameters $J_2$, $J_3$, and $\tilde{\varepsilon}$ (given $\varepsilon_1$ fixed), making the situation more complex. However, as we will show below, although the energy spectrum depends on all of the three parameters, the two groups of parameters $J_2$ and $(J_3,\tilde{\varepsilon})$ actually independently control the evolution of the BSs with varying $k$. We take this advantage to provide a quantitative interpretation of the features of the BSs observed in Fig.~\ref{numericalfig}.
\subsection{Plane-wave ansatz}
\par We now use a plane-wave ansatz to solve the two-exciton problem in the Merrifield model for $k\in K_1$. The plane-wave ansatz is a standard method to semianalytically solve the eigenvalue problem of a certain class of inhomogeneous tridiagonal matrices and has been used to treat various physical problems~\cite{Grimm,PRB2017,PRB2019}. Recently, a combination of the two-magnon Bloch states and the plane-wave ansatz is employed to completely solve the two-magnon problem in a higher-spin ferrromagnetic chain~\cite{PRBL2024}, where the evolution of the two-magnon BSs with varying wave numbers is analytically interpreted. The details of the plane-wave ansatz method is presented in Appendix~\ref{AppA} for the sake of readability.
\par Write the eigenvalue problem of the matrix $h_2(k)$ as $h_2(k)V=\lambda V$, the plane-wave ansatz gives
\begin{eqnarray}
\lambda=-4\cos p,
\end{eqnarray}
and
\begin{eqnarray}\label{Vreal}
V_j&=&\cos[(j-N/2-1)p],~j=1,2,\ldots,N/2-1,\nonumber\\
V_0&=&-\frac{2J_2}{ 4\cos p+A_k}V_1,\nonumber\\
V_{\frac{N}{2}}&=&\frac{1}{\sqrt{2}\cos p}V_{\frac{N}{2}-1},
\end{eqnarray}
where the pseudo-momentum $p$ is determined by the following transcendental equation,
\begin{eqnarray}\label{TreqK1}
\tan\frac{Np}{2}&=&f(J_2,A_k,p),\nonumber\\
 f(J_2,A_k,p)&=&\frac{2(2-J^2_2) \cos p + A_k}{2J^2_2\sin p}.
\end{eqnarray}
\par We see that $J_2$ enters the equation through $J^2_2>0$, while the onsite energies and $J_3$ enter the equation through $A_k$. The above equation should be solved on the interval $[0,\pi]$ and it may have either real or complex solutions. Although Eq.~(\ref{TreqK1}) does not admit analytical solutions and can only be solved graphically, we can still get useful information through analyzing its behaviors in some limits such as $p\to 0$ or $\pi$. Below we discuss the real and complex solutions of Eq.~(\ref{TreqK1}) in detail.
\subsection{Real solutions: Scattering states}
\begin{figure*}
\includegraphics[width=.80\textwidth]{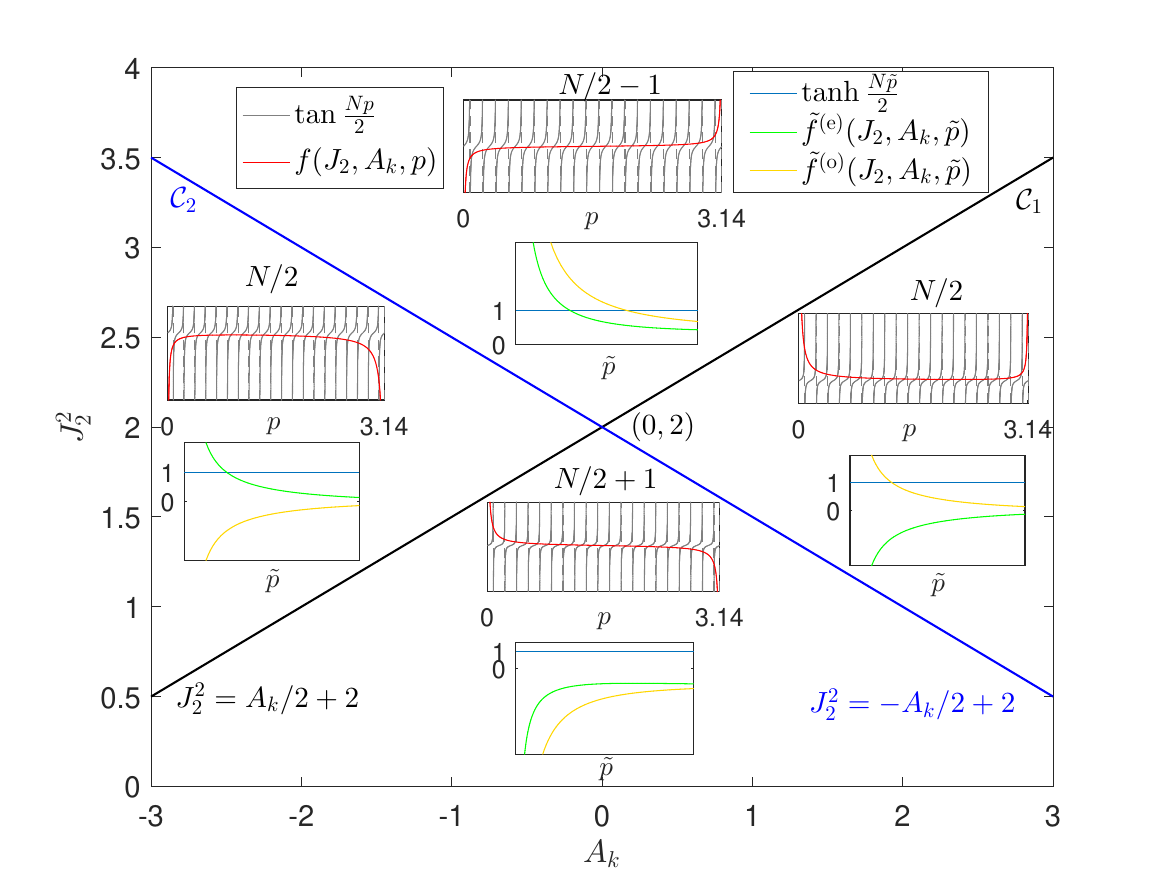}
\caption{The two straight lines defined by $J^2_2=A_k/2+2$ ($\mathcal{C}_1$, black) and $J^2_2=-A_k/2+2$ ($\mathcal{C}_2$, blue) divide the $A_k$-$J^2_2$ plane into four regions. Equation (\ref{TreqK1}) has $N/2+1$, $N/2$, $N/2$, and $N/2-1$ real solutions in the four regions (see the upper insets in each region). Correspondingly, Eq.~(\ref{tanh_eq1}) has one (no) solution above (below) $\mathcal{C}_1$; while Eq.~(\ref{tanh_eq2}) has one (no) solution above (below) $\mathcal{C}_2$ (see the lower insets in each region).}
\label{Fig2}
\end{figure*}
\par The real solutions of Eq.~(\ref{TreqK1}) give two-exciton scattering eigenstates with oscillating wave functions (\ref{Vreal}) in the Bloch space. From Eq.~(\ref{H2kCk}), the eigenenergy for each real solution $p$ is
\begin{eqnarray}
E_{2,p}(k)=2\varepsilon_1-4C_k\cos p.
\end{eqnarray}
\par It is obvious that $f(J_2,A_k,p)$ diverges as $p\to0^+$ or $p\to \pi^-$:
\begin{eqnarray}
\lim_{p\to 0^+}f(J_2,A_k,p)=\mathrm{sgn}(-J^2_2+A_k/2+2)\infty,\nonumber\\
\lim_{p\to \pi^-}f(J_2,A_k,p)=\mathrm{sgn}(J^2_2+A_k/2-2)\infty.
\end{eqnarray}
As a result, the two straight lines
\begin{eqnarray}
\mathcal{C}_1:&~&J^2_2=A_k/2+2,\nonumber\\
\mathcal{C}_2:&~&J^2_2=-A_k/2+2,\nonumber
\end{eqnarray}
divide the $A_k$-$J^2_2$ plane into four regions (Fig.~\ref{Fig2}) and they intersect at $(0,2)$. In each of the four regions we plot the graphs of both $\tan\frac{Np}{2}$ and $ f(J_2,A_k,p)$ on $[0,\pi]$ (insets in Fig.~\ref{Fig2}). These graphs clearly show that in the lower region below both $\mathcal{C}_1$ and $\mathcal{C}_2$ all the $N/2+1$ solutions of  Eq.~(\ref{TreqK1}) are real. In the right (left) region there are $N/2$ real solutions since the two functions do not intersect on the rightmost (leftmost) interval $\left[\frac{(N-1)\pi}{N},\pi\right]$ ($[0,\frac{\pi}{N}]$). Interestingly, in the upper region above $\mathcal{C}_1$ and $\mathcal{C}_2$ there are $N/2-1$ real solutions since the two functions do not intersect on both of the two intervals. Consequently, there are 0, 1, 1, and 2 complex solutions in the above four regions, respectively.
\par It should be mentioned that Knoester and Spano studied the two-photon absorption spectrum of a molecular chain that is equivalent to the Merrifield model with $J_3=0$ by using Green's function method~\cite{Spano1995PRL}. They derive an analytical expression for the spectrum, based on which a ``phase diagram" similar to Fig.~\ref{Fig2} is mapped out through counting the number of additional peaks, without relating them to two-exciton BSs.
\subsection{Complex solutions: Bound states}
\par We now look for complex solutions of Eq.~(\ref{TreqK1}) in the upper, right, and left regions of Fig.~\ref{Fig2}. To ensure the reality of the eigenenergies, a complex solution must have the form $p=m\pi+i\tilde{p}$, where $m\in \mathbb{Z}$ and $\tilde{p}\geq0$ is real~\cite{PRBL2024}. Using the relations $\cos p=(-1)^m\cosh\tilde{p}$ and $\sin p=i (-1)^m\sinh\tilde{p}$, we get
\begin{eqnarray}\label{tanh_eq1}
\tanh\frac{N\tilde{p}}{2}&=&\tilde{f}^{(\mathrm{e})}(J_2,A_k,\tilde{p}),\nonumber\\
\tilde{f}^{(\mathrm{e})}(J_2,A_k,\tilde{p})&=&\frac{2(J^2_2-2) \cosh \tilde{p} -A_k}{2J^2_2\sinh \tilde{p}},
\end{eqnarray}
for even $m$, and
\begin{eqnarray}\label{tanh_eq2}
\tanh\frac{N\tilde{p}}{2}&=&\tilde{f}^{(\mathrm{o})}(J_2,A_k,\tilde{p}),\nonumber\\
\tilde{f}^{(\mathrm{o})}(J_2,A_k,\tilde{p})&=&\frac{2(J^2_2-2) \cosh \tilde{p} + A_k}{2J^2_2\sinh \tilde{p}},
\end{eqnarray}
for odd $m$.
\par From the limiting behaviors of $\tilde{f}^{(\mathrm{e/o})}$,
\begin{eqnarray}
\lim_{\tilde{p}\to\infty}\tilde{f}^{(\mathrm{e/o})}(J_2,A_k,\tilde{p}) &=& (J^2_2-2)/J^2_2<1,\nonumber\\
\lim_{\tilde{p}\to0}\tilde{f}^{(\mathrm{e})}(J_2,A_k,\tilde{p})&=&  \mathrm{sgn}(J^2_2-A_k/2-2)\infty,\nonumber\\
\lim_{\tilde{p}\to0}\tilde{f}^{(\mathrm{o})}(J_2,A_k,\tilde{p})&=& \mathrm{sgn}(J^2_2+A_k/2-2)\infty,
\end{eqnarray}
we see that Eq.~(\ref{tanh_eq1}) [Eq.~(\ref{tanh_eq2})] has one solution in the left and upper (right and upper) regions of Fig.~\ref{Fig2} (see the insets), which is consistent with the above analysis based on the real solutions of Eq.~(\ref{TreqK1}).
\par The eigenenergy of a BS is
\begin{eqnarray}
E_{2,\tilde{p}}(k)=2\varepsilon_1-4C_k (-1)^m\cosh\tilde{p}.
\end{eqnarray}
It is now clear that a BS with odd (even) $m$ must lie above (below) the continuum since $\cosh\tilde{p}\geq 1\geq \cos p$ .
\par Let us first discuss the case of even $m$. The corresponding wave functions read
\begin{eqnarray}
V_j&=&\cosh[(j-N/2-1)\tilde{p}],~j=1,2,\ldots,N/2-1,\nonumber\\
V_0&=&RV_1,\nonumber\\
V_{\frac{N}{2}}&=&\frac{1}{\sqrt{2}\cosh \tilde{p}}V_{\frac{N}{2}-1},
\end{eqnarray}
where
\begin{eqnarray}\label{Revenm}
R&\equiv&-\frac{2J_2}{ 4\cosh \tilde{p}+A_k}.
\end{eqnarray}
It is evident that $V_1>V_2>\cdots>V_{N/2}$, so that the eigenstate is an f BS (ee BS) if $|R|>1$ ($0<|R|<1$). It is thus desirable to analytically determine in Fig.~\ref{Fig2} the two curves defined by $R=\pm1$. To this end, we consider the large $N$ limit~\cite{PRBL2024}. When $N\to\infty$, we can use the approximation $\tanh N\tilde{p}/2\approx 1$, Eq.~(\ref{tanh_eq1}) then becomes a simple quadratic equation satisfied by $e^{\tilde{p}}$:
\begin{eqnarray}\label{tanh_eq1app}
g^{(+)}(e^{\tilde{p}})&=&0,
\end{eqnarray}
where
\begin{eqnarray}\label{gplus}
 g^{(+)}(x)&\equiv& 2x^2+A_kx-2(J^2_2-1).
\end{eqnarray}
A combination of Eqs.~(\ref{Revenm}) and (\ref{tanh_eq1app}) leads to
\begin{eqnarray}\label{RpJ}
R&=&-\frac{e^{\tilde{p}}}{J_2},
\end{eqnarray}
which is always negative for $J_2>0$.
\begin{figure}
\includegraphics[width=.52\textwidth]{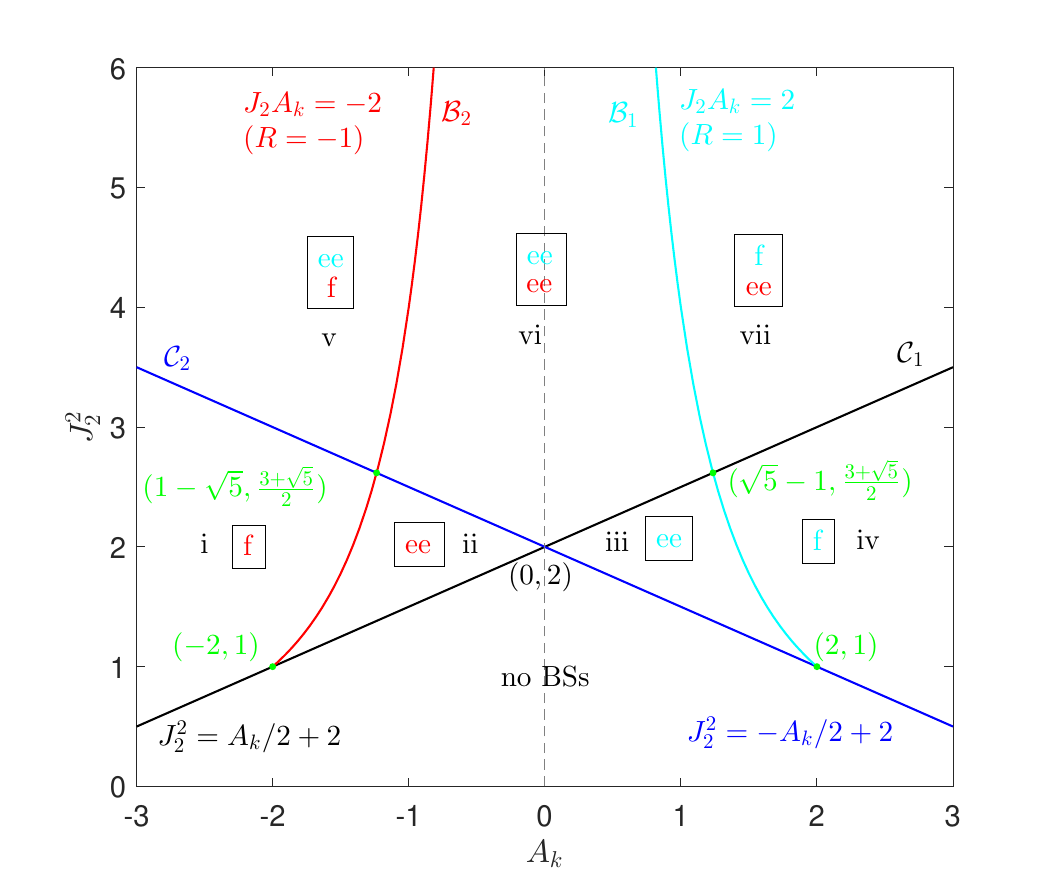}
\caption{The two curves $\mathcal{B}_{1,2}:~J_2A_k=\pm2$ with $R=\pm1$ divide the three regions in Fig.~\ref{Fig2} into seven subregions i-vii, in which the occurrence of the two types of two-exciton bound states are indicated (also summarized in Table I). The intersections of the curves are also shown.}
\label{Fig3}
\end{figure}
\par In the left and upper regions of Fig.~\ref{Fig2} we have $g^{(+)}(1)=-2(J^2_2-2-A_k/2)<0$, so $g^{(+)}(x)=0$ must have a real solution that is greater than 1, which is also the physical solution (the other solution less than 1 is obviously unphysical),
\begin{eqnarray}\label{evensol}
\tilde{x}\equiv e^{\tilde{p}}=\frac{1}{4}\left[-A_k+ \sqrt{A_k^2+16(J^2_2-1)}\right].
\end{eqnarray}
The eigenenergy can then be expressed in terms of $\tilde{x}$ as
\begin{eqnarray}
E_{2,\tilde{p}}(k)=2\varepsilon_1-2C_k (-1)^m(\tilde{x}+\tilde{x}^{-1}).
\end{eqnarray}
From Eqs.~(\ref{RpJ}) and (\ref{evensol}), the curve defined by $R=-1$ (recall that $R<0$ for $J_2>0$) satisfies
\begin{eqnarray}
J_2A_k=-2,
\end{eqnarray}
see the red curve $\mathcal{B}_2$ in Fig.~\ref{Fig3}, which intersects $\mathcal{C}_1$ and $\mathcal{C}_2$ at $(-2,1)$ and $(1-\sqrt{5},\frac{3+\sqrt{5}}{2})$, respectively. A caution is in order here: it can be checked that the point $(-\sqrt{5}-1,\frac{3-\sqrt{5}}{2})$ is also an intersection of $\mathcal{C}_1$ and $J_2A_k=-2$. However, a close look at Eq.~(\ref{evensol}) shows that $R=-1$ holds only for $J^2_2\geq 1/2$, indicating that the sub-branch of $J_2A_k=-2$ starting with $(-\sqrt{5}-1,\frac{3-\sqrt{5}}{2})$ is actually unphysical.
\begin{table}
\begin{center}
\begin{tabular}{ |c|c|c|  }
 \hline
  & even $m$ & odd $m$  \\
   \hline
 i &  f   &    \\
 ii &  ee   &    \\
iii &    & ee   \\
iv &    & f   \\
v &  f  & ee   \\
vi &  ee  & ee   \\
vii & ee   & f   \\
   \hline
\end{tabular}
\caption{Emergence of the two types of two-exciton BSs in each region of Fig.~\ref{Fig3}.}
\end{center}
\end{table}
\par Similar analysis can be applied to the case of odd $m$. In the large $N$ limit Eq.~(\ref{tanh_eq2}) becomes
\begin{eqnarray}\label{tanh_eq2app}
g^{(-)}(e^{\tilde{p}})&=&0,\nonumber\\
 g^{(-)}(x)&\equiv& 2x^2-A_kx-2(J^2_2-1).
\end{eqnarray}
In the right and upper regions of Fig.~\ref{Fig2} we have $g^{(-)}(1)=-2(J^2_2+A_k/2-2)<0$, giving the physical solution
\begin{eqnarray}
 e^{\tilde{p}}=\frac{1}{4}\left[A_k+ \sqrt{A_k^2+16(J^2_2-1)}\right],
\end{eqnarray}
and hence
\begin{eqnarray}
R&=&\frac{e^{\tilde{p}}}{J_2}>0.
\end{eqnarray}
The curve defined by $R=1$ is given by $J_2A_k=2$, i.e., the cyan curve $\mathcal{B}_1$ in Fig.~\ref{Fig3}, which intersects $\mathcal{C}_2$ and $\mathcal{C}_1$ at $(2,1)$ and $(\sqrt{5}-1,\frac{3+\sqrt{5}}{2})$, respectively.
\par Now the two curves $\mathcal{B}_1$ and $\mathcal{B}_2$ divide the original three regions (left, right, and upper) in Fig.~\ref{Fig2} into seven subregions i-vii as shown in Fig.~\ref{Fig3}. The occurrence of BSs in each region is summarized in Table I.
\par From Fig.~\ref{Fig3} we can immediately see that the necessary condition for the existence of an ee BS is $J_2>1$, while the two types of BSs cannot occur simultaneously unless $J_2>\sqrt{2}$. As we will see below, the ``phase diagram" given by Fig.~\ref{Fig3} not only serves as the basis for the identification of BSs in a fixed $k$-subspace but also provide us an opportunity to investigate the evolution of the BSs with varying $k$ observed in Fig.~\ref{numericalfig}.
\section{Evolution of the bound states with varying wave number}\label{SecIV}
\par We now discuss the evolution of the two types of BSs with varying wave number $k$ under a given set of parameters $(J_2,J_3,\tilde{\varepsilon})$. Note that $J_2$ and $(J_3,\tilde{\varepsilon})$ govern the evolution independently since $A_k=A_k(J_3,\tilde{\varepsilon},k)$ depends only on the latter. We can thus fix the value of $J_2$ and see how the abscissa $A_k$ varies in Fig.~\ref{Fig3} as $k$ increases from $-\pi$ to $0$. It is obvious that
\begin{eqnarray}
\lim_{k\to 0^-}A_k&=&\tilde{\varepsilon}+2J_3,\nonumber\\
\lim_{k\to (-\pi)^+}A_k&=&\mathrm{sgn}(\tilde{\varepsilon}-2J_3)\infty.
\end{eqnarray}
To see how $A_k$ changes on the  open interval $(-\pi,0)$, we calculate the partial derivative
\begin{eqnarray}
\frac{\partial A_k}{\partial k}=\frac{1}{2C^2_k}\sin\frac{k}{2}[\tilde{\varepsilon}-2J_3(\cos k+2)].
\end{eqnarray}
\begin{figure}
\includegraphics[width=.52\textwidth]{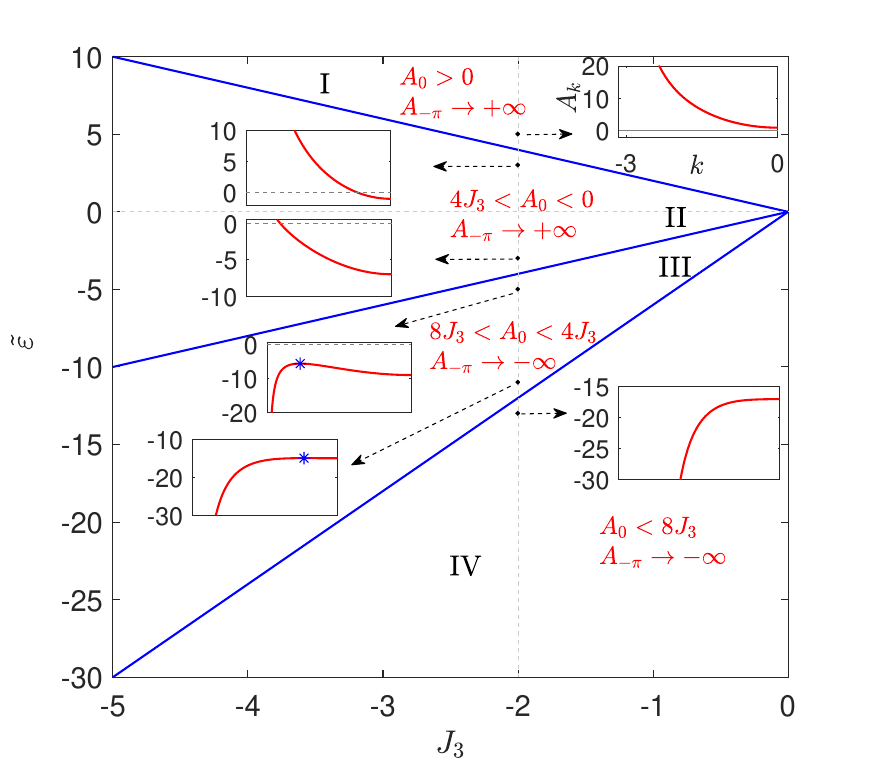}
\caption{The left half of the $J_3$-$\tilde{\varepsilon}$ plane (with $J_3<0$) is divided by the three half-lines $\tilde{\varepsilon}=-2J_3$, $\tilde{\varepsilon}=2J_3$, and $\tilde{\varepsilon}=6J_3$ into four regions. The insets in each region show the quantity $A_k(J_3,\tilde{\varepsilon},k)$ as a function of $k$ on $[-\pi,0]$ for several representative points with fixed $J_3=-2$ and varying $\tilde{\varepsilon}$. The limiting behaviors of $A_k$ as $k\to 0^+$ and $k\to (-\pi)^+$ are also listed.}
\label{Fig4}
\end{figure}
On $(-\pi,0)$ we have $\sin\frac{k}{2}<0$ and $\tilde{\varepsilon}-2J_3<\tilde{\varepsilon}-2J_3(\cos k+2)<\tilde{\varepsilon}-6J_3$. The above observations suggest us to divide the $J_3$-$\tilde{\varepsilon}$ half-plane (with $J_3<0$) into four regions I, II, III, and IV by the three straight lines $\tilde{\varepsilon}=-2J_3$, $\tilde{\varepsilon}=2J_3$, and $\tilde{\varepsilon}=6J_3$ (Fig.~\ref{Fig4}). The quantity $A_k(J_3,\tilde{\varepsilon},k)$ shows different behaviors in different regions, see the insets of Fig.~\ref{Fig4} for representative parameter sets with fixed $J_3=-2$. In region I (II) we have $\lim_{k\to(-\pi)^+}A_k=+\infty$ and $A_k$ monotonically decreases until it reaches $A_0>0$ ($4J_3<A_0<0$). Conversely, in region IV $A_k$ monotonically increases from $\lim_{k\to(-\pi)^+}A_k=-\infty$ to $A_0<8J_3$. In region III with $\lim_{k\to(-\pi)^+}A_k=-\infty$ and  $8J_3<A_0<4J_3$ there exists a wave number
\begin{eqnarray}
k^*=-\arccos(\tilde{\varepsilon}/2J_3-2),
\end{eqnarray}
at which $A_k$ reaches a maximum
\begin{eqnarray}
A_{k^*}=4\sqrt{2}J_3\sqrt{\tilde{\varepsilon}/2J_3-1}.
\end{eqnarray}
\par When $J_2$ is given, the evolution of the two-exciton BSs with increasing $k$ can then be read off by investigating the horizontal motion of a representative point (RP) of $A_k$ in Fig.~\ref{Fig3}. The trajectory of the representative point may (if it does) intersect $\mathcal{C}_{1,2}$ and $\mathcal{B}_{1,2}$ at $A_{\mathcal{C}_{1,2}}=\pm 2(J^2_2-2)$ and $A_{\mathcal{B}_{1,2}}=\pm 2/J_2$, respectively. To find out the wave numbers corresponding to these intersections, we use Eq.~(\ref{Ak}) to inversely solve $k$ in terms of $A_k$, giving
\begin{figure}
\includegraphics[width=.52\textwidth]{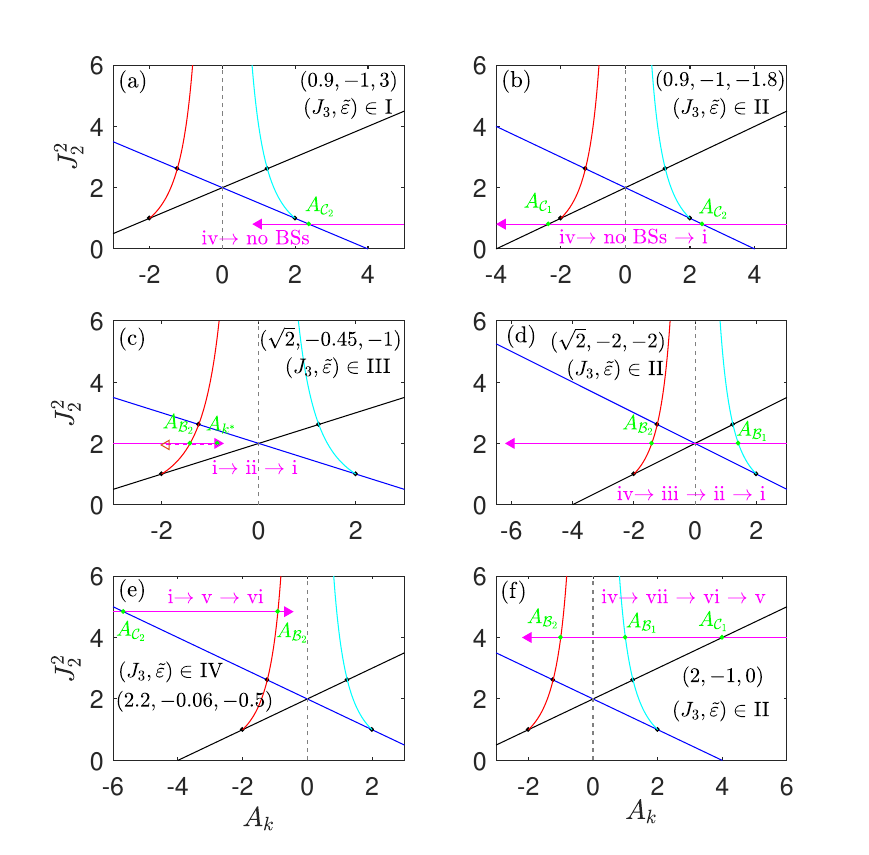}
\caption{Trajectory of the representative point of $A_k$ in the $A_k$-$J^2_2$ plane as $k$ increases from $-\pi$ to $0$ (pink). The arrow has abscissa $A_0$ [or $A_{k^*}$ in (c)] and indicates the direction of the motion of the point. The parameters are chosen in accordance with those in Fig.~\ref{numericalfig}. }
\label{intpret}
\end{figure}
\begin{eqnarray}
k^{(\pm )}&=&\chi^{(\pm )}(A_k),\nonumber\\
\chi^{(\pm )}(x)&\equiv & -2\arccos\frac{x\pm\sqrt{x^2-16J_3(\tilde{\varepsilon}-2J_3)}}{8J_3}.
\end{eqnarray}
It is possible that only one or both of the two wave numbers $k^{(\pm )}$ are physically allowed.
\par We are now in a position to interpret the observed features of the BSs in Fig.~\ref{numericalfig}. We plot in Fig.~\ref{intpret} the trajectories of the RP of $A_k$ for the six sets of parameters used in Fig.~\ref{numericalfig}. In Fig.~\ref{intpret}(a), the RP moves from region iv supporting a high-lying f BS to the region without BSs at $k_{\mathcal{C}_2}=\chi^{(-)}(A_{\mathcal{C}_2})\approx -1.0733$, which quantitatively explains the evolution of the f BS in Fig.~\ref{numericalfig}(a). In Fig.~\ref{intpret}(b) the RP keeps moving to the left and enters region i at $k_{\mathcal{C}_1}=\chi^{(-)}(A_{\mathcal{C}_1})\approx -1.6741$, where a low-lying f BS emerges. The pair of parameters $(J_3,\tilde{\varepsilon})$ used in Fig.~\ref{intpret}(c) lie in region III of Fig.~\ref{Fig4}. The low-lying f BS first changes to the ee-type at $k^{(1)}_{\mathcal{B}_2}=\chi^{(+)}(A_{\mathcal{B}_2})\approx -2.9843$ and the RP reaches the rightmost position $A_{k^*}=-0.8485$ at $k^*=-2.6657$, after which it turns back and the state returns to an f BS at $k^{(2)}_{\mathcal{B}_2}=\chi^{(-)}(A_{\mathcal{B}_2})\approx -1.5708$. In Fig.~\ref{intpret}(d) the RP passes through regions iv, iii, ii, and i consecutively and intersects $\mathcal{B}_1$ and $\mathcal{B}_2$ at $k_{\mathcal{B}_1}=\chi^{(-)}(A_{\mathcal{B}_1})\approx -2.2761$ and $k_{\mathcal{B}_2}=\chi^{(-)}(A_{\mathcal{B}_2})\approx -1.8642$, respectively. In particular, the high-lying ee BS suddenly becomes a low-lying ee BS after the RP hits $A_k=0$ at $k=-2\pi/3$. The remaining two cases [Fig.~\ref{intpret}(e) and (f)] can be interpreted in a similar way.
\par Thus, for any given set of parameters we can immediately read off the properties of the two-exciton BSs in the whole Brillouin zone by investigating the trajectory of the RP of $A_k$ in Fig.~\ref{Fig3}. In this sense, we have achieved a complete solution of the Merrifield exciton model in the two-exciton subspace.
\section{Conclusions}\label{SecV}
\par In this work, we have provided a complete and rigorous solution to the two-exciton problem in the finite-size Merrifield exciton model, a fundamental framework for understanding multi-exciton phenomena in organic molecular aggregates. By constructing a set of exact two-exciton Bloch states and employing a plane-wave ansatz to solve the resulting inhomogeneous tridiagonal matrices, we have analytically derived the full energy spectrum and the explicit forms of the wave functions for both scattering and bound states.
\par Our main results are threefold. First, we rigorously proved that at most two distinct types of two-exciton bound states can exist, i.e., the f BS where two excitons predominantly reside on a single molecular site, and the ee BS where they occupy two nearest-neighbor sites. These bound states, when present, are always either the lowest or the highest eigenstates in a given total wave-number sector. Second, we mapped out the complete parameter phase diagram in the space of $(J_2,J_3,\tilde{\varepsilon})$ (Figs.~\ref{Fig3} and \ref{Fig4}), identifying the precise conditions under which zero, one, or two bound states emerge, and distinguishing their types. In particular, we showed that an ee BS requires $J_2>1$, while the coexistence of both types necessitates $J_2>\sqrt{2}$. Third, and most importantly, we provided a quantitative and analytical interpretation of the evolution of these bound states with varying wave number $k$. By analyzing the trajectory of a representative point for the quantity $A_k$ in the phase diagram, we were able to explain the complex transitions between f-type and ee-type behavior, including the reentrant transitions and the merging of bound states into the scattering continuum, as observed in numerical simulations.
\par  The closed-form expressions of the eigenenergies and wave functions for finite-size systems can serve as valuable benchmarks for future theoretical and computational studies of exciton dynamics, nonlinear optical responses, and singlet fission processes in molecular crystals. We will address these topics in future works.

\noindent{\bf Acknowledgments:}
This work was supported by the National Key Research and Development Program of China under Grant No. 2021YFA1400803 and by the Innovation Program for Quantum Science and Technology under Grant No. 2023ZD0300700.

\appendix
\section{Plane-wave ansatz}\label{AppA}
\par Both $h_2(k)$ and $h'_2(k)$ can be incorporated into an inhomogeneous tridiagonal matrix
\begin{eqnarray}\label{hnn}
(d)_{(n+1)\times (n+1)}=\left(
      \begin{array}{cccccccc}
        a &  b_1 &   & &    &    &   &   \\
         b_1 & 0 &  b &  &   &   &   &   \\
          &  b & 0 &  b  & &   &   &   \\
          &   &  b &  0 &  &    &   &   \\
           &   &    &  & \ddots&   &   &   \\
          &   &   &     & &0 &  b &   \\
          &   &   &    &  &b & 0 &   b_2 \\
          &   &   &    & &  &  b_2  & 0 \\
      \end{array}
    \right).
\end{eqnarray}
\par For $k\in K_1$, we have
\begin{eqnarray}
n&=&\frac{N}{2},~a=A_k,~b_1= 2J_2,~b=-2,~b_2=\sqrt{2}b,\nonumber\\
\end{eqnarray}
and for $k\in K'_1$ we have
\begin{eqnarray}
n&=&\frac{N}{2}-1,~a=A_k, ~b_1= 2J_2,~b=-2,~b_2=b.\nonumber\\
\end{eqnarray}
The eigenvalue problem of $d$, i.e., $d\vec{v}=\lambda\vec{v}$, is solved by a plane-wave ansatz~\cite{PRBL2024}
\begin{eqnarray}
v_j&=&Xe^{ipj}+Ye^{-ipj},~j=2,3,\ldots,n
\end{eqnarray}
where $X$ and $Y$ are $j$-independent coefficients to be determined and the pseudo-momentum $p$ is determined by the following transcendental equation~\cite{PRBL2024},
\begin{widetext}
\begin{eqnarray}\label{Treq}
\frac{\tan np}{\sin p}=\frac{b^2_1( 2b^2-b^2_2)\cos p -b^2_2[ab -( 2b^2-b^2_1)\cos p]}{(b^2_2-2b^2)\cos p[( 2b^2-b^2_1)\cos p-ab ]+b^2_2b^2_1\sin^2p}.
\end{eqnarray}
\end{widetext}
The above equation has either real or complex solutions and can only be solved graphically. For each allowed $p$, the eigenvalue $\lambda$ is given by
\begin{eqnarray}
\lambda=2b\cos p,
\end{eqnarray}
and the (unnormalized) wave function reads~\cite{PRB2024_1}
\begin{eqnarray}
v_j=\cos[(j-N/2-1)p],~(j=2,3,\ldots,N/2)
\end{eqnarray}
The two end components are given by
\begin{eqnarray}
v_1=\frac{b_1}{\lambda-a},~~v_{n+1}=\frac{b_2}{\lambda}v_n.
\end{eqnarray}

\end{document}